# Coherent electron transport through an azobenzene molecule: A light-driven molecular switch

C. Zhang[a], M.-H. Du[a] and H.-P. Cheng[a]*, X.-G. Zhang[b], A.E. Roitberg[c] and J.L. Krause[c]

[a]Department of Physics and Quantum Theory Project, University of Florida, [b] Oak Ridge National Laboratory, [c]Department of Chemistry and Qauntum Theory Project, University of Florida

## Abstract

We apply a first-principles computational approach to study a light-sensitive molecular switch. The molecule that comprises the switch can convert between a *trans* and a *cis* configuration upon photo-excitation. We find that the conductance of the two isomers vary dramatically, which suggests that this system has potential application as a molecular device. A detailed analysis of the band structure of the metal leads and the local density of states of the system reveals the mechanism of the switch.

* Corresponding author

The next generation of electronic devices will undoubtly be constructed of molecules, or with features of molecular size. The results of recent experimental [1-3] and theoretical studies [4-11] predict a brilliant future for molecular electronics. Among various efforts and activities, several schemes have been proposed to design and construct molecular switches [3, 8-11]. The basic idea is to find molecules that have two or more distinct states with vastly different conductance. Switching between *on* and *off* states can be performed by applying an external bias or by using a scanning tunneling microscope (STM) tip to manipulate the system [3, 8-11]. These methods are not ideal since the former may interfere greatly with the function of a nano-size circuit, and the latter imposes severe limitations in device applications. In addition, these methods are relatively slow. Alternative methods, such as those based on fast, light-driven processes, are therefore highly desirable. The photo-sensitive azobenzene molecule discussed in this work is one such candidate for an opto-electronic device.

In the ground electronic state, the azobenzene has two energetically equivalent conformations [12-14], *trans* and *cis*, which are separated by a barrier of approximately 1.0 eV. The *trans* structure is 1.98 Å longer than the *cis*. The molecule can switch reversibly from one structure to another under photo-excitation, with the structural change occurring on an electronically excited state [14,15].

Previous work has shown that a nano-scale opto-mechanical device based on the conformation change of the azobenzene polymer can be designed. In the experiment [15], light with a wavelength of 365 nm transforms azobenzene units from *trans* structures to *cis* structures. As a result, the polymer contracts and retracts an atomic force microscope (AFM) tip, thus converting the energy of electromagnetic radiation to mechanical work. A second pulse of light with a wavelength of 420 nm reverses the *cis* back to the *trans* configuration.

In this paper, we report investigations of transport in the molecular wire, Au-S-azobenzene-S-Au, which consists of a single azobenzene molecule, which has been functionalized by replacing a hydrogen atom with a $CH_2S$ group to provide a good contact with gold, and two semi-infinite gold leads. Note that the name of the modified molecule is technically bis-(4-methanethiol-pheyl) diazene or BMTPD. Since one can add more azobeneze units to elongate the polymer, we simply follow Ref. 14-15 and refer the molecular system as "azobenze". As shown in Fig. 1, the entire system is divided into three regions, the left lead, the switch and the right lead. The switch includes two layers of gold from each lead. The structures of the leads are chosen to be the same as in Ref. 16, which corresponds to a gold (111) crystal lattice.

Also shown in Fig.1 is one possible realization of the azobenzene molecular switch. The left Au lead is held fixed and the right gold rod is contained within a conducting carbon nano-tube (CNT). Changing the length of the molecule via photo-excitation results in a change in current in the CNT.

The electronic structure of the switch region is determined by treating it as a cluster. Full quantum calculations, based on density functional theory (DFT), are used to optimize the structure of a cluster consisting of the molecule, the sulfur atoms and two rigid layers of gold. The optimization is performed with two different choices of the inter-lead distance, $d$. For *trans*, the calculated optimal $d$ is 18.16 Å (Fig.1a), and for *cis*, the optimal $d$ is 16.23 Å (Fig.1b).

The next step is to calculate the conductance of the two optimized structures. In this study, we focus on the linear-response regime. Within this limit, the entire system is in equilibrium, and the Fermi surface is a constant. Due to charge screening by electrons in the metal, the two leads, as well as the switch, are charge-neutral. The conductance is calculated by the Landauer formula [17], $\sigma = 2e^2 T(E_F)/h$, where $T(E_F)$ is the transmission function at the Fermi energy. The effective potentials are calculated by first-principles density functional theory (DFT). For the wire system depicted in Fig.1, the system is, in principle, infinite, and periodic boundary conditions are not appropriate. However, due to screening, the Kohn-Sham potentials in the two semi-infinite leads are not affected by the switch. As a result, the potentials can be computed by separate calculations of a one-dimensional gold wire with periodic boundary conditions [5-7].

With Green's function techniques, the effect of the two leads on the switch can be absorbed into self-energy terms, which are calculated using the same method as in Ref. 16. The transmission function $T(E_F)$ is evaluated by the Caroli formula [18,19]. The density of states (DOS) in the switch region can be calculated from the trace of the retarded Green's function, which includes both scattering states and localized states. The DOS can then be projected onto any desired subsystem. Note that the sub-system can be chosen as a single atom or a group of atoms.

The calculated transmission as a function of energy and the DOS for the *trans* and *cis* structures, with optimal inter-lead distance *d* are shown in Fig. 2. Unlike the DOS of an isolated molecule, Fig. 2 displays a series of peaks superimposed on a broad background. These features originate from a combination of localized states from the molecule and a nearly continuous energy band of the two leads. At some energies, the DOS shows peaks, but the total transmission is very small. Notice, for example, peaks *a*, and *b*, for the *trans* isomer, and peaks *a'*, and *b'* for the *cis* isomer. These peaks are indications of localized states, which can be seen more clearly in the local density of states.

As depicted in Fig. 3, the DOS of the switch region is decomposed into three sub-systems, the two Au layers, the two S atoms, and the azobenzene molecule. For the *trans* isomer, peak *a* is mainly localized on the contacts, namely, the Au layers and S atoms, while peak *b* is localized mainly on the molecule. For the *cis* isomer, peak *a'* is localized mainly in the Au layers, while peak *b'* has high density on the molecule. The DOS on the S atoms is very small in the *cis* isomer as compared to the same peak in the *trans* isomer.

These comparisons indicate that the *trans* and *cis* configurations have quite different contacts with the leads. Furthermore, Fig. 3 indicates clearly that the broad background in the DOS derives primarily from the Au layers.

Figure 3 shows that the Fermi energy of the system lies between the highest occupied molecule orbital (HOMO) and the lowest unoccupied molecule orbital (LUMO), slightly closer in energy to the HOMO. Compared to the isolated molecule, the peaks in the local DOS projected on the molecule indicate that the molecular orbitals are perturbed by the presence of the leads and contribute to scattering states near the Fermi energy. Since we focus here on the zero-bias case, the conductance is determined completely by the transmission at the Fermi energy, which is dominated by the tail of the electron distribution tunneling through the perturbed HOMO. Note that the local DOS of both the Au layers and the S atoms peak near the *trans* HOMO energy, but that the magnitude of these peaks are minimal for the transmission function in Fig.2. For the *trans* isomer, tunneling through the perturbed HOMO leads to a very broad and intense peak in the transmission function, but a sharp and smaller peak for the *cis* isomer (Fig.2, lower panel). As a result, at the Fermi energy, tunneling is more rapid through the *trans* structure than through the *cis* structure, although the DOS of the *cis* isomer is higher than that of the *trans* isomer. The calculated zero-bias conductance for the *trans* structure is $0.21 \times 10^{-4}\, \Omega^{-1}$, which indicates a moderately good conductor. In contrast, the conductance for the *cis* structure is nearly two orders of magnitude lower.

To understand the tunneling process at the Fermi energy, Fig. 4 depicts the local DOS at the Fermi energy for both the *trans* and the *cis* isomers. In this figure, we decompose the DOS into 7 components; the two left Au layers, the left S atom, the left benzene ring with a $CH_2$ group, the double-bonded $N_2$, the right benzene ring with a $CH_2$ group, the right S atom, and the two right Au layers. As can be seen in the figure, the local DOS for the two molecular conformations differs significantly. For the *cis* isomer, the local DOS is much higher on the Au layers and the two benzene rings than on the S atoms and the N double bond. The nearly zero DOS on the right S atom, and the oscillation of the DOS along different sites suggest that the scattering states are localized primarily on the benzene rings and the gold leads. Reflection of the scattering wave occurs at sites of low DOS (on S and N). Both localization and reflection contribute significantly to the resistance of the device. Compared to the *cis* isomer, the local DOS for the *trans* isomer displays a smoothly varying curve suggesting that localization and reflection at the N double bond and the two S atoms contribute much less to resistance than in the *cis* configuration.

In the azobenzene/gold system, the extended scattering states allow facile electron-tunneling through the molecule. This result can also interpreted from the viewpoint of molecular structure. In the *cis* configuration, due to the different orientations of the two benzene rings, orbitals on different rings have different configurations. Consequently, when electrons tunnel through the azo unit of the *cis* configuration, they are scattered more than they are in the *trans* configuration.

In conclusion, the conductance of the *cis* isomer is significantly smaller than that of the *trans*. Since the azo-benzene molecule can easily change its configuration between *trans* and *cis* via excitation with light, we suggest that this system is a viable candidate as a light-driven molecular switch. The switch will work at room temperature due to the high thermal stability of both the *trans* and the *cis* configurations of the azobenzene molecule. Compared to switches driven by an external bias field or by mechanical movement of an STM tip, the device discussed here will be much easier to control and to synthesize. Since previous experimental work has demonstrated techniques to build even more complicated devices using azobenzene molecules, we believe that the switch suggested in this paper can be fabricated in the laboratory. Further investigations are underway to optimize the parameters of this and similar molecule switches.

**Acknowledgement:**

Chun Zhang acknowledges support from the University of Florida in the form of an alumni fellowship. This work is supported by the Department of energy/Basic Science (contract DE-FG02-02ER45995). The authors acknowledge computing support from DOE/NERSC and the Oak Ridge Supercomputer Center.

Fig.1 (a) Physical models of an azobenzene switch in contact with two gold leads: *trans* isomer (upper) and *cis* isomer (lower); (b) Schematic of one possible realization of a light-driven switch. The left Au lead is held fixed and the right gold rod is contained within a conducting carbon nano-tube (CNT). Changing the length of the molecule via photo-excitation results in a change in current in the CNT.

Fig. 2 Density of states (DOS) projected onto the switch region (upper panel) and transmission as a function of energy (lower panel). The solid lines represent the *trans* configuration and dotted lines the *cis*. Intense peaks in the DOS are truncated to improve visibility.

Fig. 3 Local DOS of the *trans* and *cis* configurations decomposed into contributions on the Au layers (upper panel), S atoms (middle panel) and the azo-benzene molecule (lower panel). The solid lines represent the *trans* configuration and dotted lines the *cis*.

Fig. 4 Local DOS at the Fermi energy projected on seven sites in the switch. From 1-7, or left to right: Au lead, S atom, benzene ring, double-bonded $N_2$, second benzene ring, S atom and Au lead. The filled squares represent the *trans* configuration and the open circles the *cis*.

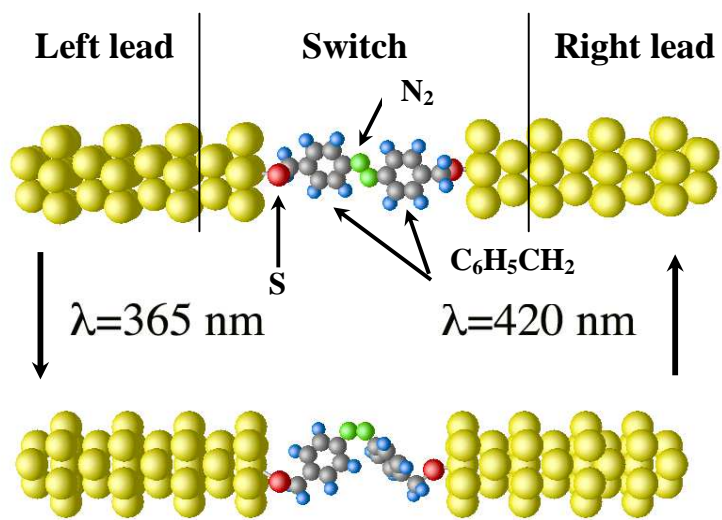

(a) Physical models

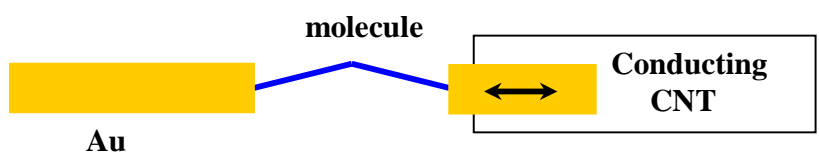

(b) Schematic of realization

Fig.1

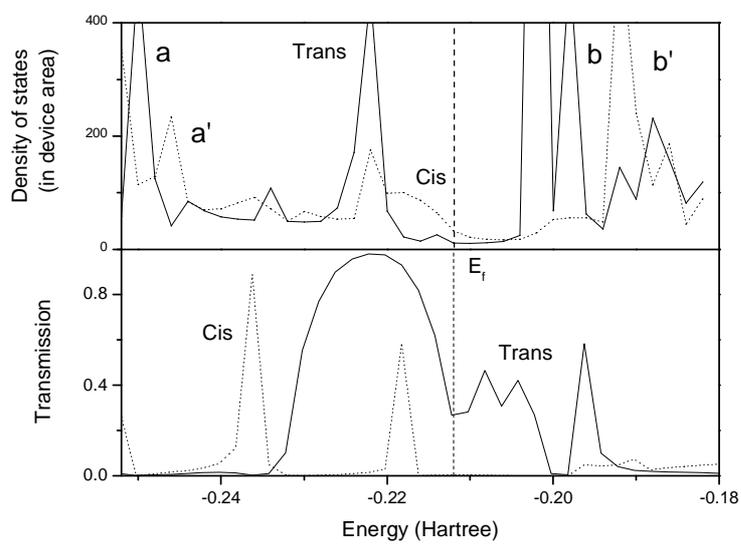

Fig. 2

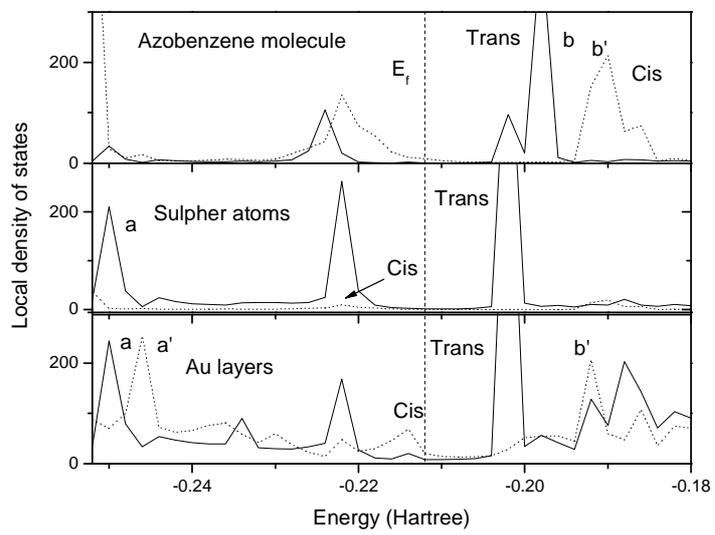

Fig.3

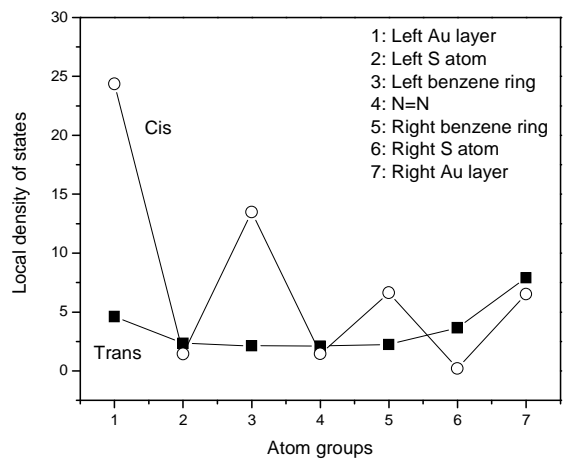

Fig.4